 \documentstyle[prl,twocolumn,aps]{revtex} 
\begin{document} 
\draft
\title{		Average Entropy of a Subsystem
}
\author{	Siddhartha Sen\cite{author}
}
\address{
		School of Mathematics,
		University of Dublin,
		Trinity College,
		Dublin,
		Ireland
}
\date{		\today
}
\maketitle

\begin{abstract}

It was recently conjectured by D. Page that if a quantum system of Hilbert
space dimension $nm$ is in a random pure state then the average entropy
of a subsystem of dimension $m$ where
$
	m \leq n
$
is
$
	S_{mn} = \sum^{mn}_{k=n+1}(1/k) - (m-1)/2n .
$
In this letter this conjecture is proved.

\end{abstract}


	In a recent letter Page \cite{page} considered
	a system $AB$ with Hilbert space dimension $mn$.
	The system consisted of two subsystems $A$ and $B$
	of dimensions $m$ and $n$ respectively.
	Page calculated the average

\begin{eqnarray*}
			S_{mn} =  <S_A> 
\end{eqnarray*}
	of the entropy $S_A$ over all pure states
$
		\rho= \mid\Psi><\Psi\mid
$
	of the total system where
$
		S_{A} = - \mbox{ Tr } \rho_{A} \ln \rho_{A}
$
	and $\rho_{A}$, the density matrix of subsystem A, is obtained by
	taking the partial trace of the full density
	matrix $\rho$ over the other subsystem, that is,
$
		\rho_{A} = \mbox{Tr}_B \rho.
$

	The average was defined with respect to the unitary invariant
	Haar measure on the space of unitary vectors $\mid\Psi>$ in
	the $mn$ dimensional Hilbert space of the total system.
	The quantity $(\ln m - S_{mn})$ was used to define the
	average information of the subsystem A.
	It is a measure of the information regarding the fact that
	the entire system is a pure state that is contained in the
	subsystem $m$.  Using earlier work \cite{lubkin,lloyd}
	in this area, Page was led to consider the probability
	distribution of the eigenvalues of $\rho_A$ for the
	random pure states $\rho$ of the entire system.
	He used

\begin{eqnarray*}
&&	P(p_{1}, \ldots, p_{m}) \, dp_{1} \ldots dp_{m}
\\&&	=
	N \delta(1-\sum_{i=1}^{m}p_{i})
\;
	\prod_{1\leq i < j \leq m}(p_i - p_j )^2
\;
	\prod^{m}_{k=1}p^{n-m}_{k}dp^{}_{k}
\end{eqnarray*}
	where $p_{i}$ was an eigenvalue of $\rho_A$ and the
	normalisation constant for this probability distribution
	was given only implicitly by the requirement
	that the total probability integrated to unity.
	Page then showed that the average

\begin{eqnarray*}
	S_{mn}
&	=
&	- \int \biggl( \sum^m_{i=1} p_i \ln p_i \biggr)
\,
	P(p_1, \ldots, p_m) \, dp_1 \ldots dp_m
\\&	=
&	\psi(mn+1)
	-
	\frac{\int\bigl(\sum^m_{i=1} q_i \ln q_i \bigr)\,Q\,dq_1 \ldots dq_m}
		{ mn \int \! Q \, dq_{1} \ldots dq_{m} } 
\end{eqnarray*}
	where $q_i = r p_i$ for $i = 1, \ldots, m$, $r$ is positive
	\cite{page}, and

\begin{eqnarray*}
	\psi(mn+1) = -C + \sum_{k=1}^{mn}\frac{1}{k}
\,	,
\end{eqnarray*}
	$C$ being Euler's constant, and

\begin{eqnarray*}
	Q(q_{1}, ..., q_{m}) dq_{1}... dq_{m}
	=
\!\!\!	\prod_{1 \leq i < j \leq m} \!\!\!\! (q_i - q_j)^2
	\prod^{m}_{i=1} e^{-q_i} q_i^{n-m} dq_i .
\end{eqnarray*}
	On the basis of evaluating $S_{mn}$ for $m =$ 2, 3, 4, 5
	using {\sc mathematica} 2.0,
	Page conjectured that the exact result for $S_{mn}$ was

\begin{eqnarray*}
	S_{mn}=\sum^{mn}_{k=n+1}\frac{1}{k}-\frac{(m-1)}{2n}
\,	,
\end{eqnarray*}
	but was not able to prove that this was the case.
	In this letter, we will prove this conjecture.

	We first observe that the van der Monde determinant
	defined by
\[
	\Delta(q_1, \ldots, q_m)
\,\,	\equiv
	\prod_{i \leq i < j \leq m}(q_{i}-q_{j})
						\nonumber
\]
	may be written
\[
	\Delta(q_1, \ldots, q_m)
	=
	\left|\matrix
{
	1	& \cdots	&	1
\cr
	q_1	& \cdots	&	q_m
\cr
	\vdots	& \ddots	&	\vdots
\cr
     q_1^{m-1}	& \cdots	&	q_m^{m-1}
\cr
}	\right| 
\,	.					\nonumber
\]
	We next observe that $\Delta(q_1,...,q_m)$ can be written as

\begin{equation}
	\Delta(q_1, \ldots, q_m)
	=
	\left| \matrix
{
	p_{0}(q_{1})	& \cdots	& p_{0}(q_{m})
\cr
	p_{1}(q_{1})	& \cdots	& p_{1}(q_{m})
\cr
	\vdots		& \ddots	& \vdots
\cr
	p_{m-1}(q_{1})	& \cdots	& p_{m-1}(q_{m})
\cr
}	\right|
\label{polydet}
\end{equation}
	for any set of polynomials $p_{k}(q)$,
$
	k = 0, \ldots, m-1,
$
	which have the property,
$
	p_{0}(q)=1,
$
	and

\begin{eqnarray*}
	p_{k}(q) = q^{k} + C_{k-1}q^{k-1}+ \cdots + C_{0},
\quad
	k=1, \ldots, m-1
\,
	.
\end{eqnarray*}
	This immediately follows from the fact that the value of a
	determinant does not change if the multiple of any one row
	is added to a different row.

	We now choose polynomials $p^{\alpha}_{k}(q)$ judiciously.
	We introduce orthogonal polynomials
$
		p_{k}^{\alpha}(q)
$
	with the properties:

\begin{enumerate}
\item
$	p_{k}^{\alpha}(q)
\;	=
\;	q^{k} + C_{k-1}q^{k-1} + \cdots + C_{0}^{\alpha} ,
\quad
	p_{0}^{\alpha}(q)=1$.
\item
	$\int^{\infty}_{0} dq e^{-q} q^{\alpha}
	p_{k_1}^{\alpha}(q) p_{k_2}^{\alpha}(q)
	=
	h^{\alpha}_{k_1}\delta_{k_1 k_2},
\,\,\,\,					\alpha = n - m.
$
\end{enumerate}
	Polynomials with these properties are well known.
	They are the generalised Laguerre polynomials defined by
\cite{prudnikov}

\begin{eqnarray*}
	p_{k}^{\alpha}(q)
	=
	\frac{e^{q}}{q^{\alpha}} (-1)^{k}
	\frac{d^{k}}{dq^{k}} \bigl( e^{-q}q^{k+\alpha} \bigr)
\,
	.
\end{eqnarray*}
	We also note, for later use, that \cite{prudnikov}

\begin{equation}
	p_{k}^{\alpha}(q)
	=
	\sum_{r=0}^{k}
%
%
	{k \choose r}	
	(-1)^{r}\frac{\Gamma(k+\alpha+1)}{\Gamma(k+\alpha-r+1)}q^{k-r}
						\label{genpoly}
\end{equation}

\begin{equation}
	\int^\infty_0 \!dq\,e^{-q}q^\alpha p_{k_1}^\alpha(q)p_{k_2}^\alpha(q)
	=
	\Gamma(k_1 + 1) \Gamma(k_1 + \alpha + 1) \delta_{k_{1}k_{2}}
						\label{twopoly}
\end{equation}

\begin{equation}
	\int^{\infty}_{0} \! dq \, q^{a-1}e^{-q}p_{k}^{b}(q)
	=
	(1-a+b)_k \Gamma(a)(-1)^k
						\label{onepoly}
\end{equation}
	recalling that, $(1-a+b)_k = (1-a+b)(1-a+b+1) \ldots (1-a+b+k-1)$.
	Writing $\Delta(q_1, \ldots, q_m)$ in terms of $p^{\alpha}_{k}(q)$
	as in Eq.\ \ref{polydet}
	and using the orthogonal property of these polynomials
	it immediately follows that:

\begin{eqnarray*}
	S_{mn}
&	=
&	\psi(mn+1)
\\&&	- \frac{1}{mn} \sum_{k=0}^{m-1} \int^{\infty}_{0}
	\frac{ e^{-q}\,(q \ln q)\,q^{n-m} \bigl(p_k^{m-n}(q)\bigr)^2 dq}
	{\Gamma(k + 1) \Gamma(k + 1 + n - m)}
\,	.
\end{eqnarray*}
We thus need to evaluate the integral

\begin{eqnarray*}
	I_{nm}^{k}
&	=
&	\int^{\infty}_0 (q\ln q) \, q^{n-m}
\,
	\bigl( p_k^{m-n}(q) \bigr)^2 \, e^{-q} \, dq
\,	.
\end{eqnarray*}
We first introduce

\begin{eqnarray*}
	J^{k}(\alpha)
&	=
&	\int^{\infty}_0 q^{\alpha+1}
\,
	\bigl( p_k^\alpha(q) \bigr)^2 \, e^{-q} \, dq
\,	.
\end{eqnarray*}
	This integral is easily evaluated.  We have

\begin{eqnarray}
	J^{k}(\alpha)
&	=
&	\Gamma(k+1)\Gamma(k+\alpha+2) + k^2 \Gamma(k)\Gamma(k+\alpha+1)
						\label{jayalpha}
\end{eqnarray}
	and we now note that

\begin{eqnarray*}
	I_{nm}^{k}
&	=
&	\left[ \frac{dJ^{k}(\alpha)}{d\alpha}
	-
	2 \int^{\infty}_0 dq \, q^{\alpha +1}e^{-q}
\,
	p^{\alpha}_{k} \, \frac{dp_k^\alpha}{d\alpha} \right]_{\alpha = n-m}
\,	.
\end{eqnarray*}
	Evaluating these two terms using Eqs.~(\ref{genpoly}),
	(\ref{twopoly}), (\ref{onepoly}), and (\ref{jayalpha}) we find

\begin{eqnarray}
	S_{mn}
&	=
&	\psi(mn+1)					\nonumber
\\&&	-						\nonumber
	\frac{1}{mn}\sum^{m-1}_{k=0}\bigl[1+(1+2k+n-m)\psi(k+n-m+1)\bigr]
\\&&	+
	\frac{2}{mn}\sum^{m-1}_{k=0}\sum^{k}_{r=0}	\nonumber
	{k \choose r} (-1)^{k+r}
	\frac{\Gamma(k+n-m+1)}{\Gamma(k+n-m-r+1)}	\nonumber
\\&&
	\times \bigr[\psi(k+n-m+1) - \psi(k+n-m-r+1)\bigr] \nonumber
\\&&
	\times \frac{(r-k-1)_k\Gamma(k+n-m-r+2)}{\Gamma(k+1)\Gamma(k+n-m+1)}
							\label{twoterm}
\end{eqnarray}
	where we use the fact that
$
		\psi(z)=\frac{1}{\Gamma(z)}\frac{d\Gamma(z)}{dz}.
$
	We now observe that

\begin{eqnarray}
&&	\psi(mn+1)					\nonumber
\\&&	-						\nonumber
	\frac{1}{mn}\sum_{k=0}^{m-1}[1+(1+2k+n-m)\psi(1+k+n-m)]
\\&&	=
	\sum_{k=n+1}^{mn}\frac{1}{k} + \frac{(m-1)}{2n}
\,	.
\end{eqnarray}
	This follows by examining the coefficient of $\frac{1}{r}$ in 

\begin{eqnarray*}
	\sum^{m-1}_{k=0} (1+2k+n-m) \psi(1+k+n-m)
\end{eqnarray*}
	and writing

\begin{eqnarray*}
	\psi(1+k+n-m) & = & -C + \sum_{r=1}^{k+n-m}\frac{1}{r}
\,	.
\end{eqnarray*}
	The third expression in Eq.~(\ref{twoterm}) above is

\begin{eqnarray}
&&	\frac{2}{mn}\sum^{m-1}_{k=0}\sum^{k}_{r=0}	\nonumber
	{k \choose r} (-1)^{k+r}
	\frac{\Gamma(k+n-m+1)}{\Gamma(k+n-m-r+1)}
\\&&							\nonumber
	\times \bigr[\psi(k+n-m+1)-\psi(k+n-m-r+1)\bigr]
\\&&							\nonumber
	\times \frac{(r-k-1)_k\Gamma(k+n-m-r+2)}{\Gamma(k+1)\Gamma(k+n-m+1)}
\\&&	=
	\frac{2}{mn} \sum_{k=0}^{m-1}			\nonumber
	{k \choose 1} (-1)^{2k+1}			\nonumber
\\&&	=
	-2 \, \frac{(m-1)}{2n}
\,	.
\end{eqnarray}
	Since $(r-k-1)_{k}=0$, for all
	$r \neq 0$ and $r \neq 1$,
	and also
	$\psi(k+n-m+1)-\psi(k+n-m-r+1)=0$ when $r=0$,
	we obtain

\begin{equation}
	S_{mn} = \sum^{mn}_{k=n+1}\frac{1}{k}-\frac{(m-1)}{2n}
\end{equation}
	as conjectured by Page.

%
This work is part of project SC/218/94 supported by Forbairt.




\begin{references}
%
\bibitem[*]{author}
		Electronic address: 
{\tt					sen\,@\,maths.tcd.ie
}
\bibitem{page}
		D. N. Page, Phys.~Rev.~Lett. {\bf 71}, 1291 (1993).
\bibitem{lubkin}
		E. Lubkin, J. Math.~Phys.~{\bf 19}, 1028 (1978).
\bibitem{lloyd}
		S. Lloyd and H. Pagels,
		Ann.~Phys.~(N. Y.) {\bf 188}, 186 (1988).
\bibitem{prudnikov}
	A. P. Prudnikov, Yu. A. Brychkov, and O. I. Marichev,
{\em	Integrals and Series Vol.2,
}
	Gordon and Breach Publishers (1988).

\end{references}
\end{document}